\documentclass{article}
\usepackage{spconf}         % ICASSP スタイル
\usepackage{amsmath,amssymb}
\usepackage{graphicx}
\usepackage{hyperref}
\usepackage{array}
\usepackage{booktabs}
\usepackage{multirow}
\usepackage{makecell}
\usepackage{tabularx}
\usepackage{siunitx}
\usepackage[most]{tcolorbox}
\usepackage{mdframed}
\usepackage{enumitem}
\usepackage[dvipsnames]{xcolor}
\usepackage{adjustbox}

\title{Hallucination Localization in Video Captioning}
\name{%
  \begin{tabular}{ccc}
    Shota Nakada & Kazuhiro Saito & Yuchi Ishikawa \\
    Hokuto Munakata & Tatsuya Komatsu & Masayoshi Kondo
  \end{tabular}
}
\address{LY Corporation}

\begin{document}
%\ninept
%
\maketitle
\begin{abstract}
We propose a novel task, hallucination localization in video captioning, which aims to identify hallucinations in video captions at the span level (i.e. individual words or phrases).
This allows for a more detailed analysis of hallucinations compared to existing sentence-level hallucination detection task.
To establish a benchmark for hallucination localization, we construct HLVC-Dataset, a carefully curated dataset created by manually annotating 1,167 video-caption pairs from VideoLLM-generated captions.
We further implement a VideoLLM-based baseline method and conduct quantitative and qualitative evaluations to benchmark current performance on hallucination localization.
\end{abstract}
\section{Introduction}
\label{sec:intro}

% Video platforms, such as Netflix and YouTube, have experienced rapid growth. 
Video streaming and sharing platforms have experienced rapid growth in recent years.
This has led to unprecedented volumes of video content.
This expansion has made automatic video understanding an important research area in computer vision.
Among the various tasks within video understanding, video captioning, which describes video content using natural language, has garnered particular attention~\cite{abdar2024review}.
Video captioning is highly valuable as it provides summaries for users and facilitates effective video content search and recommendation. 

% Recently, VideoLLMs have become widely utilized in video captioning tasks~\cite{videochat2, videochat}.
Video captioning tasks have recently made extensive use of VideoLLMs~\cite{videochat2, videochat}.
VideoLLMs are models that integrate a video encoder with a large language model (LLM)~\cite{grattafiori2024llama3, achiam2023gpt} to perform various video-language tasks, such as answering questions, describing scenes, and summarizing video content.
While VideoLLMs generate versatile and fluent captions, they inherit the hallucination problem common in LLMs, producing content that is not supported by or contradicts the input video~\cite{huang2025survey, ma2024vista}. 
Such hallucinated captions may mislead users and diminish the system's trustworthiness, particularly when captions serve as official summaries or input for downstream tasks. 
Therefore, addressing hallucination in video captioning is crucial for deploying VideoLLMs safely and reliably.

\begin{figure}[t!]
  \centering
  \includegraphics[width=7.3cm]{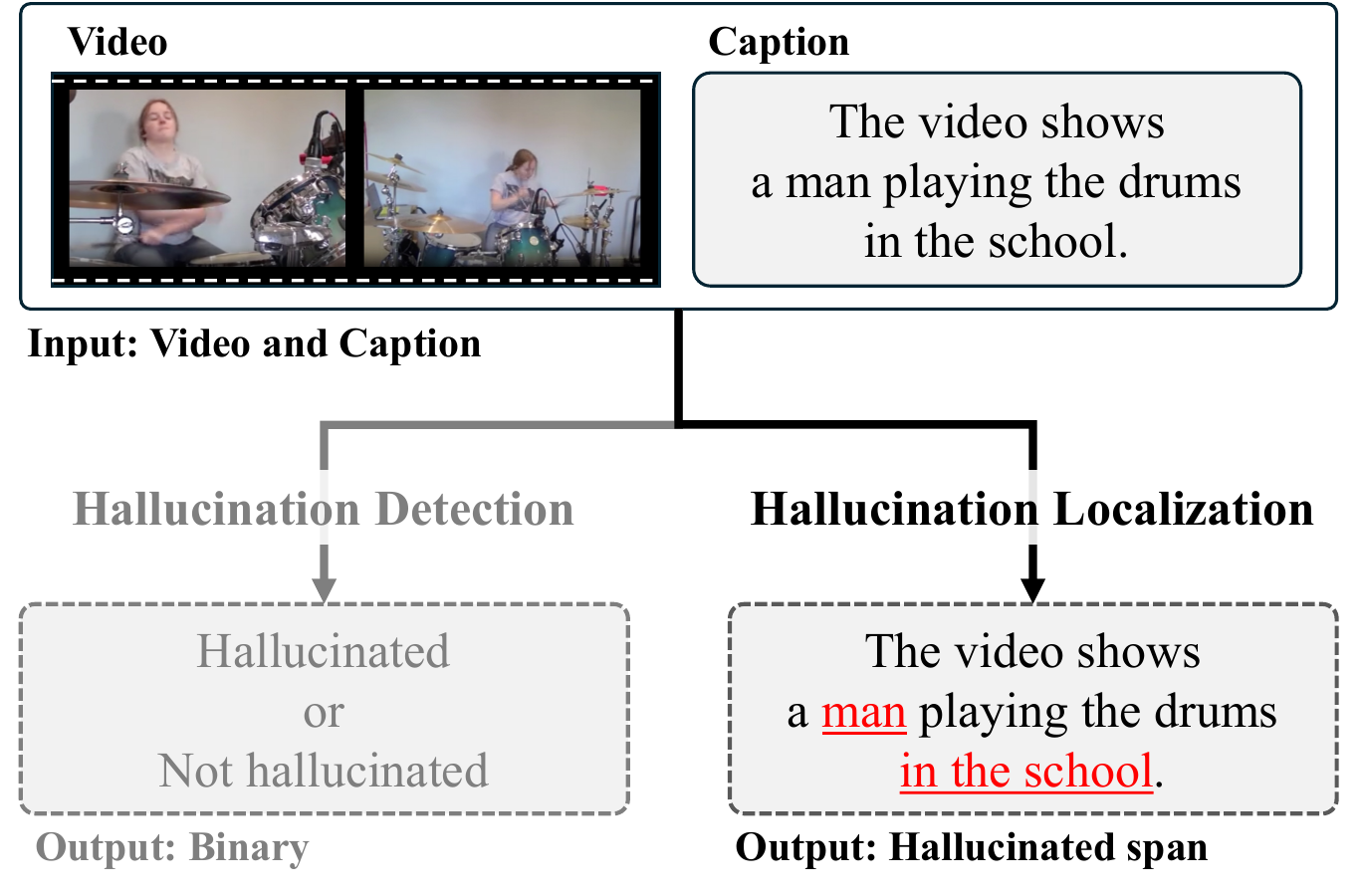}
  \vspace{-4mm}
  \caption{\textbf{Comparison between the hallucination detection and hallucination localization.} Given a video and its caption, hallucination detection classifies whether the caption contains hallucinated content. In contrast, our proposed hallucination localization identifies the text span responsible for the hallucination.}
  \label{fig:intro}
  \vspace{-4mm}
\end{figure}

Hallucination in video captioning has been actively explored by researchers~\cite{choong2024vidhal, ullah2022thinking}.
Among the various research directions on hallucination, sentence-level hallucination detection, which involves identifying incorrect captions at the sentence level, is particularly important~\cite{shi2022emscore, liu2023factvc}.
Hallucination detection in video captioning is typically formulated as a binary classification task, where the model determines whether a caption contains hallucinations.
This detection step is essential for providing feedback about caption errors to users, thereby preserving the overall reliability of video systems.

However, sentence-level hallucination detection suffers from critical limitations due to its coarse granularity. 
While existing hallucination detection methods operate at the sentence level, hallucinations in video captions typically occur at finer granularity, such as individual words or phrases. 
For example, Fig.~\ref{fig:intro} illustrates a case where the caption `\textit{The video shows a man playing the drums in the school}' is provided for a video that actually depicts a woman playing drums somewhere. 
Here, hallucinations occur only within specific spans, such as `\textit{man}' and `\textit{in the school}'.
Existing sentence-level hallucination detection overlooks these detailed errors, thereby limiting thorough analysis of caption quality. 
Moreover, providing only sentence-level warnings to users does not specify the exact source of hallucination, making feedback inadequate. 
Therefore, detailed, fine-grained feedback is critical for precise evaluation and user-oriented services.

To address these issues, we propose a novel task and a designated new dataset for hallucination localization in video captioning. 
Hallucination localization aims to precisely identify textual spans (words or phrases) within captions that contradict visual evidence from the corresponding video. 
By enabling span-level detection, our approach provides accurate feedback to users by marking only erroneous segments, thus preserving correct information. 
This fine-grained localization not only enhances caption reliability but also provides valuable guidance for model improvement and potentially increases interpretability.

We construct the HLVC-Dataset by generating captions using multiple VideoLLMs on videos selected from existing datasets such as MSR-VTT~\cite{xu2016msr} and FAVD-Bench~\cite{shen2023favd}, and manually annotating each hallucinated span. 
The resulting dataset comprises 1,167 video-caption pairs, each containing at least one hallucinated segment. 
Additionally, we propose a VideoLLM-based baseline model for hallucination localization.  
This baseline utilizes instruction-tuned VideoLLMs to generate hallucinated spans as output.
We conduct extensive experiments using five different VideoLLMs and evaluate their performance quantitatively and qualitatively.

\section{Proposed Task}
To enable a more fine-grained analysis beyond sentence-level hallucination, we propose the task of hallucination localization. In this section, we describe the definition of the task and the evaluation metrics.

%-------------------------------------------------
\subsection{Task Definition}
\label{sec:task_definition}

Let the evaluation set contain $M$ video–caption pairs.  
For the $j$-th sample ($1\!\le\! j\!\le\! M$) we denote the video by
$v^{(j)}$ and its caption by
\(
\mathbf{x}^{(j)}=( x^{(j)}_{1},\dots,x^{(j)}_{n_{j}}).
\)
The objective of hallucination localization is
to decide, for every token $x^{(j)}_{i}$, whether it is grounded
in the visual evidence of $v^{(j)}$.  
We model a hallucination localization system as a function
$f(\mathbf{x},v)$ and write
\vspace{-2mm}
\[
\hat{\mathbf{y}}^{(j)} = f(\mathbf{x}^{(j)}, v^{(j)}) = (\hat{y}^{(j)}_{1}, \dots, \hat{y}^{(j)}_{n_{j}}).
\]
\noindent Here, each predicted token label is
\vspace{-3mm}
\[
\hat{y}^{(j)}_{i} =
  \begin{cases}
    1 & \text{if }x^{(j)}_{i}\text{ is hallucinated},\\[4pt]
    0 & \text{otherwise}.
  \end{cases}
  \vspace{-2mm}
\]
Contiguous indices with $\hat{y
}^{(j)}_{i}=1$ constitute
hallucination spans.  
This task requires fine-grained language and vision alignment along with precise error tagging.

%-------------------------------------------------
\subsection{Evaluation Metrics}
\label{sec:evaluation_metrics}
For each sample we assume an oracle label sequence  
\(\mathbf{y}^{(j)}=( y^{(j)}_{1},\dots,y^{(j)}_{n_{j}})\).
%Inspired by the evaluation protocols in Grammatical Error Detection~\cite{bell-etal-2019-context},  
%we adopt precision, recall, accuracy, and F$_{0.5}$ based metrics to assess system performance.  
%Specifically, we evaluate predictions at both the token and span levels:
We define evaluation metrics at both the token-level and the span-level.

\noindent \textbf{Token-level evaluation:}  
  We directly compare predicted hallucination labels $\hat{\mathbf{y}}^{(j)}$  
  against oracle labels $\mathbf{y}^{(j)}$ for each token.  
  Inspired by the evaluation protocols in Grammatical Error Detection~\cite{bell-etal-2019-context},  
we adopt precision, recall, accuracy, and F$_{0.5}$, thereby measuring fine-grained correctness of hallucination localization.
  %Precision, recall, accuracy, and F$_{0.5}$ are computed over the set of tokens,  
  %thereby measuring fine-grained correctness of hallucination localization.

\noindent \textbf{Span-level evaluation:}
  We further evaluate predictions at the hallucination span level.
Following the protocol introduced in Named Entity Recognition~\cite{segura-bedmar-etal-2013-semeval}, we define Accuracy under two different settings: \emph{strict} and \emph{partial}.
In the \emph{strict} setting, a predicted hallucination span is considered correct only if it exactly matches the oracle-labeled hallucination span.
In the \emph{partial} setting, a predicted span is considered correct if it overlaps with an oracle-labeled hallucination span by at least one token.
This dual evaluation captures both exact boundary identification and more flexible overlap-based correctness.

  % \item \textbf{Sentence-level evaluation:}  
  %At the sentence level, a sample is considered hallucinated if at least one token in  
  %$\mathbf{x}^{(j)}$ is labeled as hallucination in the oracle.  
  %Likewise, system predictions are binarized at the sentence level by checking  
  %whether any token is predicted as hallucination.  
  %Precision, recall, accuracy, and F$_{0.5}$ are then computed over sentences,  
  %reflecting the system’s ability to judge whether a caption as a whole  
  %contains hallucination.

\begin{figure*}[t!]
  \centering
  \includegraphics[width=15.7cm]{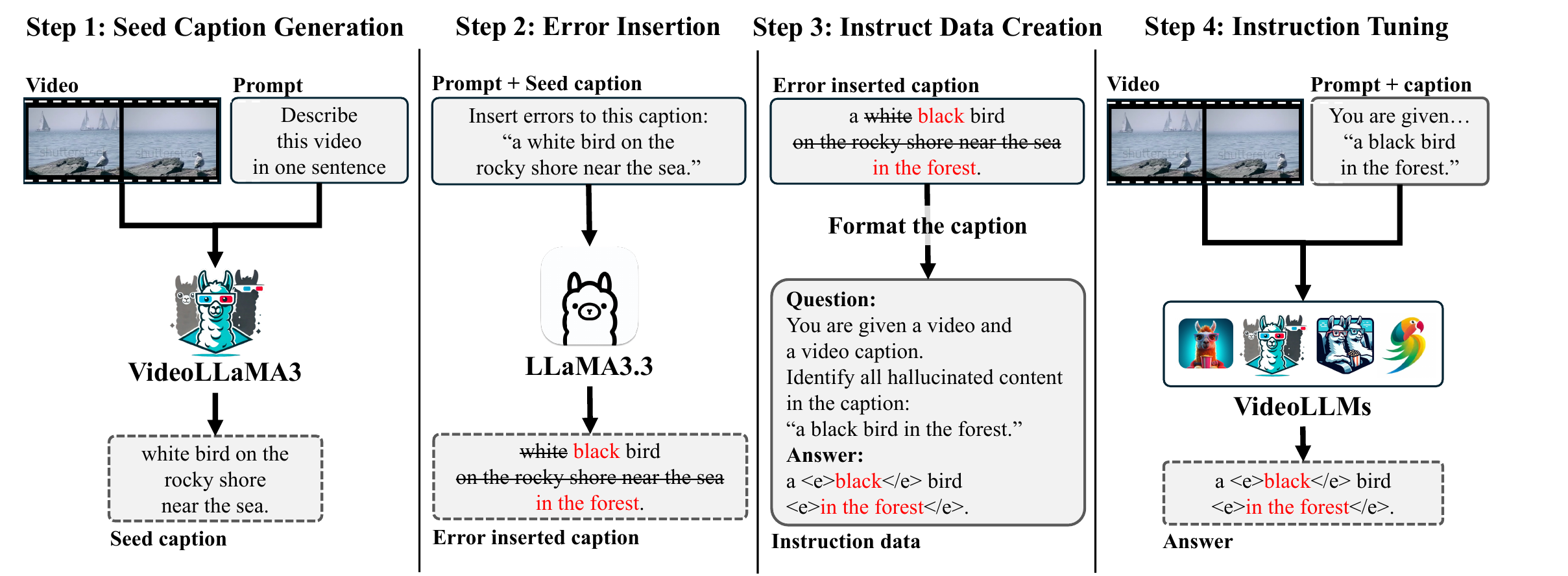}
  \vspace{-4mm}
  \caption{\textbf{Overview of instruction tuning framework.} The procedure is as follows: Step 1 generates seed captions using an existing VideoLLM. Step 2 automatically inserts errors into these seed captions using an LLM (LLaMA3.3). Step 3 formats the error-inserted captions as instruction data for VideoLLMs. Step 4 performs instruction tuning on VideoLLMs specifically for hallucination localization, enabling the tuned model to output hallucinated spans in the input video captions.
}
    \vspace{-4mm}
  \label{fig:method}
\end{figure*}

\section{Dataset: HLVC-Dataset}
In this section, we present HLVC‑Dataset, a new benchmark expressly designed for Hallucination Localization in Video Captioning (HLVC). 

\noindent \textbf{Video dataset selection.}
We collected videos from existing video datasets.
We used MSR‑VTT~\cite{xu2016msr} and FAVD‑Bench~\cite{shen2023favd} as our sources.
MSR‑VTT is one of the most widely used corpora for video captioning research and includes diverse, open‑domain footage.
FAVD‑Bench is a video dataset designed for tasks that take audio information into account and offers high audiovisual diversity.
We extracted 1,000 clips from each dataset, gathering 2,000 videos in total.

\noindent \textbf{Video caption generation.}
Video captions are automatically generated using existing VideoLLMs.
We select VideoLLaMA~\cite{videollama} and VideoChat~\cite{videochat}, the most recent models available when our annotation began.
By providing each video together with the prompt “Describe this video in one sentence.” to the VideoLLMs, we obtain its caption.
Applying both VideoLLMs to the 2,000 videos yields 4,000 video–caption pairs in total.

\noindent \textbf{Annotation Protocol.}
We annotate hallucinations in video captions through a three-stage procedure. The first stage involves a caption-level decision, where we determine whether the caption contains any hallucination. A binary label is assigned: captions are labeled as 1 if at least one hallucination is present, and 0 otherwise. Hallucinations are further categorized into two types. The first type, termed invented, refers to content that cannot be verified from the video, such as describing the setting as “in the school” when the environment is not discernible. The second type, termed contradictory, refers to content that explicitly conflicts with the visual evidence, for instance, describing “a man” when the person in the video is in fact a girl.
In the second stage, for captions labeled as 1, each hallucinated span is explicitly marked by enclosing it in error tags. This process specifies both the number and the precise location of hallucinations within the text.
Finally, in the third stage, each marked span is minimally revised through either substitution or deletion. These edits aim to eliminate the hallucinations while preserving the overall fluency and meaning of the caption as much as possible.

\begin{table}[ht]
  \centering
  \caption{\textbf{Hallucination localization performance on HLVC-Dataset.} 
We report token-level (Accuracy, Precision, Recall, and F$_{0.5}$) and span-level (strict and partial) performance.
N/A indicates cases where evaluation was not possible due to model outputs not conforming to the required format.}

    \label{tab:performance_comparison}
  \setlength{\tabcolsep}{4pt} %
  \scalebox{0.85}{
  \begin{tabular}{lcccccccc}
    \toprule[1.2pt]
    Model & \multicolumn{4}{c}{Token-level} & \multicolumn{2}{c}{Span-level} \\
    \cmidrule(lr){2-5} \cmidrule(lr){6-7}
           & Acc & P & R & F$_{0.5}$ 
           & Acc(strict) & Acc(partial) \\
    \midrule[0.5pt]
    \multicolumn{7}{l}{\textbf{Zero-Shot}} \\
    \midrule[0.5pt]
    VideoChat      & N/A  & N/A & N/A & N/A & N/A  & N/A \\
    VideoChat2     & 69.9 & 21.2 & 15.8 & 19.8 & 15.2 & 24.4 \\
    VideoLLaMA     & N/A  & N/A & N/A & N/A & N/A  & N/A \\
    VideoLLaMA2.1  & 79.1 & 64.7 & 1.4 & 6.3 & 2.0 & 3.1 \\
    VideoLLaMA3    & 78.5 & 31.0 & 1.6 & 6.7 & 1.3 & 2.1 \\
    \midrule[0.5pt]
    \multicolumn{7}{l}{\textbf{Ours (Instruction Tuning)}} \\
    \midrule[0.5pt]
    VideoChat      & 78.0 & 35.5 & 5.5 & 17.1 & 4.7 & 13.2 \\
    VideoChat2     & 80.4 & 63.9 & 15.3 & 39.1 & 14.2 & 27.5 \\
    VideoLLaMA     & 79.6 & 56.8 & 12.8 & 33.7 & 12.9 & 28.3 \\
    VideoLLaMA2.1  & 81.0 & 64.8 & 21.4 & 46.1 & 18.4 & 34.5 \\
    VideoLLaMA3    & \textbf{82.9} & \textbf{68.6} & \textbf{34.3} & \textbf{57.2} & \textbf{30.1} & \textbf{45.6} \\
    \bottomrule[1.2pt]
  \end{tabular}
  }
\end{table}

\section{Method}
In this section, we present the baseline method for hallucination localization. As illustrated in Fig.~\ref{fig:method}, our approach is based on instruction tuning of VideoLLM and consists of four steps in total.
\label{sec:baseline}

\noindent \textbf{Seed Caption Generation.}
The objective of this step is to produce a large corpus of video–caption pairs. 
We sample 500,000 videos from WebVid2M~\cite{bain2021webvid}, a large-scale video dataset. 
For each video, we prompt VideoLLaMA3~\cite{videollama3} with “Describe this video in one sentence.” to generate a corresponding caption. 
We then compute the video–caption similarity scores with LanguageBind~\cite{zhu2023languagebind}, retain the 10,000 highest-scoring pairs, and use them as seed data for instruction tuning.

\noindent \textbf{Error Insertion.}
The objective of this step is to generate video captions that deliberately contain hallucinations. Drawing on the FAVA framework, we implement a procedure that uses an LLM to inject synthetic errors.
We define three error categories (Entity, Relation, Invented) following FAVA~\cite{mishra2024fine}. Each error type is inserted into the seed caption with a fixed insertion probability, and we ensure that the different errors do not interfere with one another.
In our experiments, we employ LLaMA3.3-70b~\cite{grattafiori2024llama3} for error injection and set the insertion probability for each error type to 0.5.

\noindent \textbf{Instruct Data Creation.}
We convert captions containing inserted errors into an instruction-based format. Specifically, we create instruction data by concatenating an erroneous caption with the following prompt: "You are given a video and a video caption. Identify all hallucinated content in the caption:". The expected output of this instruction data is the original caption, with inserted errors enclosed within the tags. If a caption contains no errors, the instruction data output remains unchanged from the original caption.

\noindent \textbf{Instruction Tuning.}
We perform instruction tuning on pretrained VideoLLMs using the created instruction data. 
After instruction tuning, the model localizes hallucinated content within video captions by enclosing hallucinated spans with  tags. 
In our experiments, we utilize five VideoLLMs: VideoChat~\cite{videochat}, VideoChat2~\cite{videochat2}, VideoLLaMA~\cite{videollama}, VideoLLaMA2~\cite{videollama2}, and VideoLLaMA3~\cite{videollama3}. 
The instruction data format and tuning parameters generally follow default settings. Additionally, all models employ a unified LLM architecture of 7 billion parameters.
\section{Experiments}
\begin{figure}[t!]
  \centering
  \includegraphics[width=8.6cm]{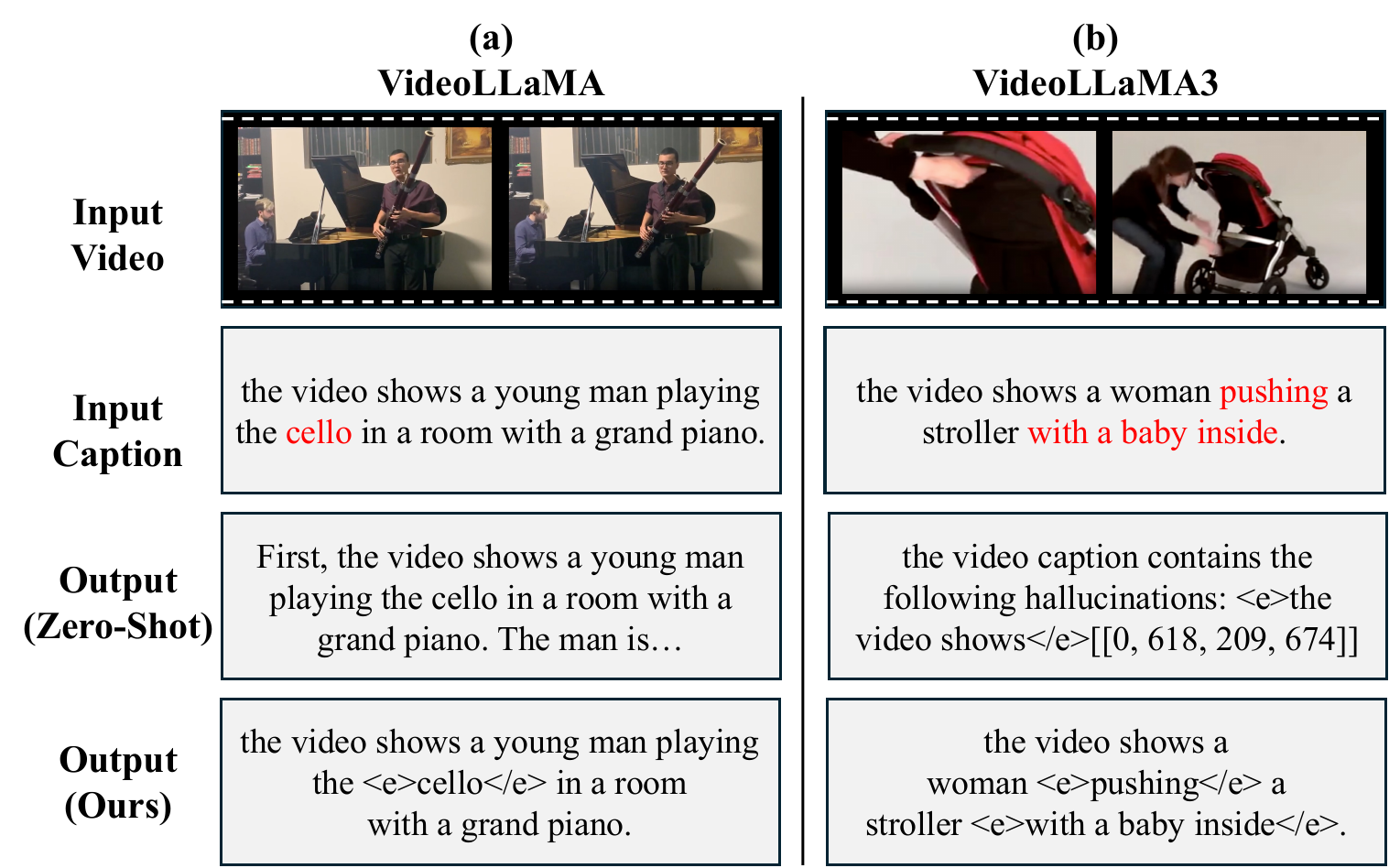}
  \vspace{-4mm}
  \caption{\textbf{Qualitative evaluation of hallucination localization.} The first column lists the input video, the second the input caption, the third the model output in the zero-shot setting, and the fourth the model output produced with our instruction-tuned method. The spans highlighted in red within the input caption indicate hallucinated spans.}
  \label{fig:quantative}
  \vspace{-4mm}
\end{figure}

\noindent \textbf{Hallucination Localization.}
To evaluate the hallucination localization capability of the models, we extract a test set comprising 500 video-caption pairs containing hallucinations from the HLVC-Dataset.
We perform evaluations under both zero-shot and instruction-tuned model, employing the same prompt format for consistency across comparisons.
As described in Section \ref{sec:evaluation_metrics}, we conducted evaluation at both the token-level and the span-level.

Table~\ref{tab:performance_comparison} summarizes the performance of the models under both evaluation settings. 
In the zero-shot scenario, models demonstrate significant limitations, frequently failing to output the correct format, and, even when they succeed, they yield relatively low performance. 
Conversely, our method considerably improves performance, highlighting its effectiveness in enhancing hallucination localization.
Notably, VideoLLaMA3 achieves an accuracy of 30.1\% at the span-level.
Additionally, there are instances where VideoChat and VideoLLaMA correctly identify their own generated hallucinations, indicating that the instruction tuning process enhances models' introspective capabilities.

\noindent \textbf{Qualitative Evaluation.}
Figure~\ref{fig:quantative} shows qualitative evaluations of hallucination localization results. 
It compares the zero-shot outputs with those of our instruction-tuned method for VideoLLaMA and VideoLLaMA3.
(a) shows the results for VideoLLaMA. Although the video depicts a man playing a "bassoon," the caption incorrectly describes him playing a cello. In the zero-shot scenario, the instruction is ignored, and the model merely describes the video content. Conversely, our method accurately localizes the hallucination span. This example demonstrates that VideoLLaMA can identify its own hallucinations, indicating the potential for self-correction in captions.
(b) provides an example of a caption containing multiple hallucinations. In the zero-shot scenario, the output is in an incorrect format, but our method correctly localizes each hallucination span. This indicates that our model can handle not only simple cases such as individual word errors but also more complex hallucinations.

\begin{figure}[t]
  \centering
  \includegraphics[width=8.6cm]{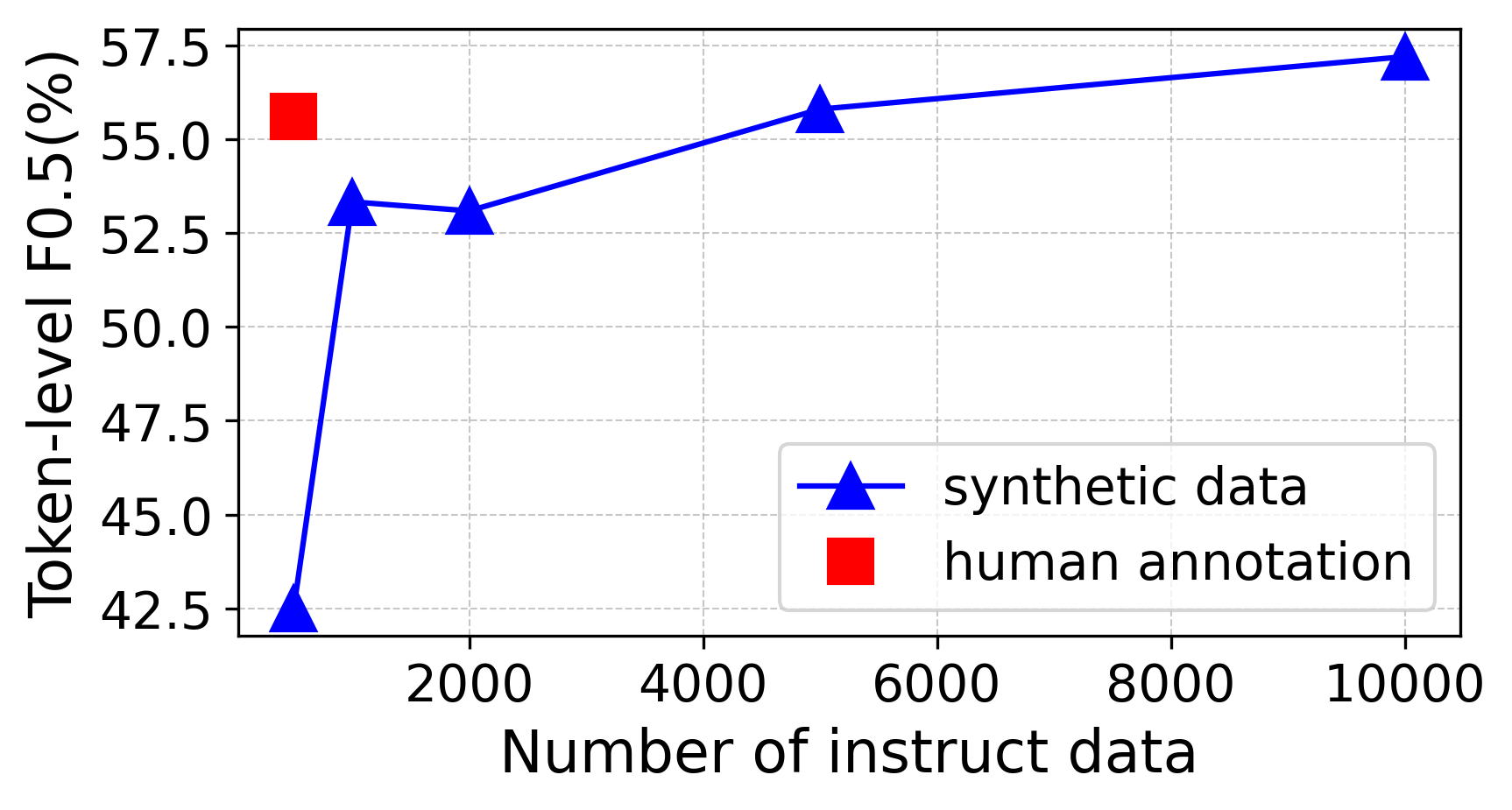}
  \vspace{-4mm}
  \caption{\textbf{Comparison of human annotation and synthetic data in instruction data}. The horizontal axis represents the number of instruction data, while the vertical axis represents the token-level F$_{0.5}$.}
  \label{fig:ablation_data_num}
  \vspace{-4mm}
\end{figure}

\noindent \textbf{Ablation Study: Human-annotation vs Synthetic Data.} 
Figure \ref{fig:ablation_data_num} shows how F$_{0.5}$ changes when using human-annotated instruction data versus synthetic data on VideoLLaMA3.
At 500 samples, human annotation yields 55.6 \%, while synthetic data only reaches 42.5 \%. 
As we increase simulated data size, F$_{0.5}$ steadily climbs, exceeding 55.8\% at around 5,000 samples. This simple trend demonstrates that, beyond a modest sample threshold, large-scale simulation can outperform manual annotation. Combining a small human-annotated set with additional synthetic examples may therefore offer an efficient path to high-quality instruction data.

\section{Conclusion}
In this paper, we introduced a novel task, hallucination localization in video captioning, which specifically identifies spans within captions containing content not grounded in the corresponding visual evidence. To facilitate research on this task, we created the HLVC-Dataset, a carefully annotated dataset consisting of 1,167 video-caption pairs with marked hallucination spans. Additionally, we proposed an instruction-tuned VideoLLM baseline designed to accurately predict hallucinated spans.
Our experiments demonstrated that instruction tuning significantly enhances the ability of VideoLLMs to localize hallucinations, achieving substantial improvements over zero-shot approaches. We further provided extensive qualitative and quantitative evaluations, illustrating our method's effectiveness and highlighting areas for potential improvement.
Future work should incorporate a broader range of error scenarios and enhance overall model reliability in video captioning applications.

\section{Acknowledgement}
We would like to express my sincere gratitude to Mr. Taichi Nishimura for his invaluable contributions to the completion of this study. In particular, his work on constructing the dataset played a crucial role in laying the foundation for this research.

\bibliographystyle{IEEEbib}
\bibliography{refs}

\end{document}